# A Review on Internet of Things (IoT), Internet of Everything (IoE) and Internet of Nano Things (IoNT)

Mahdi H. Miraz
Department of Computer Science and Software Engineering
University of Ha'il, Ha'il, Saudi Arabia
Glyndŵr University, Wrexham, United Kingdom
m.miraz@uoh.edu.sa, m.miraz@glyndwr.ac.uk

Maaruf Ali
School of Architecture, Computing and Engineering
University of East London, Docklands Campus
London, E16 2RD, United Kingdom
M.Ali3@uel.ac.uk, maaruf@ieee.org

Peter S. Excell, Rich Picking
Glyndŵr University
Wrexham, United Kingdom
p.excell@glyndwr.ac.uk, r.picking@glyndwr.ac.uk

*Abstract*—The current prominence and future promises of the Internet of Things (IoT), Internet of Everything (IoE) and Internet of Nano Things (IoNT) are extensively reviewed and a summary survey report is presented. The analysis clearly distinguishes between IoT and IoE which are wrongly considered to be the same by many people. Upon examining the current advancement in the fields of IoT, IoE and IoNT, the paper presents scenarios for the possible future expansion of their applications.

## I. INTRODUCTION

The applications and usage of the Internet are multifaceted and expanding on a daily basis. The Internet of Things (IoT), Internet of Everything (IoE) and Internet of Nano Things are new approaches for incorporating the Internet into the generality of personal, professional and societal life. This paper examines the current state of these technologies and their multi-dimensional applications by surveying the relevant literature. The paper also evaluates the various possible future applications of these technologies and foresees further developments and how these will change the way that life will be lived in the future.

## II. THE INTERNET OF THINGS (IoT)

The term 'Internet of Things' or 'Internet of Objects' has come to represent electrical or electronic devices, of varying sizes and capabilities, that are connected to the Internet. The scope of the connections is ever broadening to beyond just machine-to-machine communication (M2M). IoT devices employ a broad array of networking protocols, applications and network domains [*1*]. The rising preponderance of IoT technology is facilitated by physical objects being linked to the Internet by various types of short-range wireless technologies such as ZigBee, RFID, sensor networks and through location based technologies [*2*]. IoT will make the impact of the Internet even more pervasive, personal and intimate in daily life [*3*]. The emergence of IoT as a distinctive entity was achieved, according to the CISCO Internet Business Solutions Group (IBSG), when more inanimate objects were connected to the Internet than human users [*3*]. According to this definition, this occurred in mid-2008. This is an accelerating ongoing process especially with the rollout of CISCO's 'Planetary Skin', the Smart Grid and intelligent vehicles [*3*].

IoT devices are not currently strongly standardized in how they are connected to the Internet, apart from their networking protocols. IoT may be employed with added management and security features to link, for example, vehicle electronics, home environmental management systems, telephone networks and control of domestic utility services. The expanding scope of IoT and how it can be used to interconnect various disparate networks is shown in Fig. 1 [*3*].

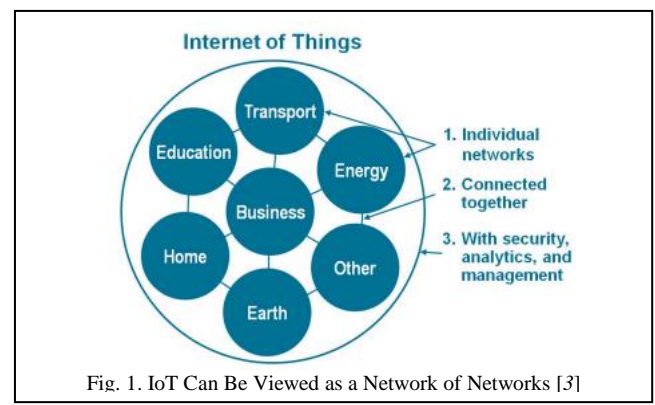

Fig. 1. IoT Can Be Viewed as a Network of Networks [*3*]

*A. Challenges and Impediments to IoT*

As with any new technology, there is usually some inertia in the pace of its uptake. Currently the largest three impediments are due to technological factors and not human resistance, these being: standardization of protocols, implementation of IPv6 and power needed to supply the sensors.

*B. Deployment of IPv6*

In February 2011 [4] the supply of IPv4 addresses held by the Internet Assigned Numbers Authority (IANA) was exhausted. The ushering in of IPv6 was critical to cover this IP address shortage, as billions of sensors will each require a unique IP address. The deployment of IPv6 will further make network management less complex with its enhanced security features and network auto-configuration capabilities.

*C. Sensor Energy*

Supplying reliable power to the sensors for a prolonged period of time is key to IoT being deployed successfully [3]. This is especially of major concern where these sensors are employed in remote and distant locations such as underground or in space or on other planets. Energy must be harvested from the environment itself as it is not feasible to change the batteries for billions of these devices. Several technologies are being pursued in order to achieve this, including solar cells, thermal generators (using the Seebeck effect), rectification of radio signals and exploitation of the energy in vibrations and other peripheral movements.

*D. Standardization*

Foremost in addressing the latest requirements for implementing IoT in terms of privacy, security and network architecture is the work of the IEEE standardization organization, especially in regard to IPv6 packet routing through heterogeneous networks [3].

### III. INTERNET OF EVERYTHING (IoE)

Both CISCO and Qualcomm have been using the term 'Internet of Everything' (IoE) [5,6]. However, Qualcomm's use of the term has been replaced by the 'Internet of Things' (IoT) by others. CISCO's usage has a more comprehensive meaning.

IoE is built upon the "four pillars" of people, data, process and things. Whereas IoT is only composed of "things", as shown in Fig. 2. IoE also extends business and industrial processes to enrich the lives of people.

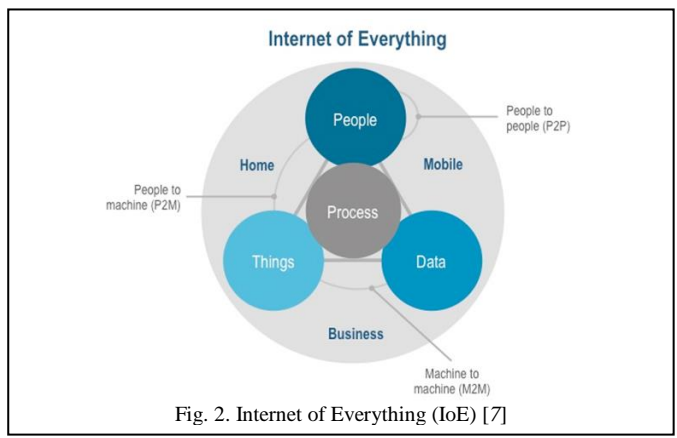

Fig. 2. Internet of Everything (IoE) [7]

The independent devices of the past are now being connected to the Internet including machine-to-machine (M2M), person-to-machine (P2M), and person-to-person (P2P) systems. This enveloping of people, processes, data and things by IoE is shown in Fig. 2 [6,7].

The Futurist at CISCO, Dave Evans, states that the issue is more about not the 'things' but the "connections among people, process, data, and things" that is at the heart of Internet of Everything and creates the 'value' [8]. Qualcomm CEO, Steve Mollenkopf, stated in 2014 that the IoT and IoE were "the same thing" [5].

According to CISCO, many organizations are going through growth waves of S-curves, as shown in Fig. 3. These IoT growth waves are leading to the eventual complete IoE [6,9]. With each successive wave of added features and greater network connectedness, this leads to the IoE with many novel opportunities as well as risks [10].

The IoE has the potential to extract and analyse real-time data from the millions of sensors connected to it and then to apply it to aid "automated and people-based processes" [11]. Other benefits include the use of IoE in helping to achieve public policy goals, environmental sustainability, economic and social goals [11].

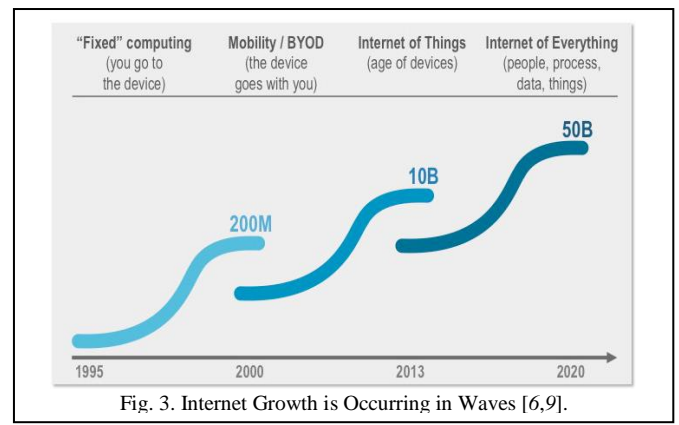

Fig. 3. Internet Growth is Occurring in Waves [6,9].

Traditional office based applications such as financial trading have now moved into the domain of the mobile platform with the use of smart phones as well as many other applications, aided by IoE [12,13]. The application of IoE is facilitated by the expansion of Cloud Computing, helping to connect "everything" online [14].

A study by CISCO in February 2013 predicted that $14.4 trillion may be exploited in the next ten years by implementing IoE with M2M, M2P and P2P [14].

Cities which in the future may be regarded as a scaled version of the IoE, will benefit the most from being connected in terms of using information intelligence to address city specific concerns. This will become more so as cities become "Smart Cities" utilizing IoE along with 'Big Data' processing. Examples include monitoring the 'health' of highways and attending to their repairs using road embedded sensors; road traffic flow control, agricultural growth monitoring, education and healthcare [15]. The future is most likely to be cities as "Smart+Connected Communities" formed using public-private partnerships to help enhance the living conditions of the citizens.

As urbanization continues to increase, predicted to be 70% by the 2050s [15], the use of IoE will become almost critical in implementing such features of the future city as the Smart Grid and automation of traffic planning and control. IoE is also forming a foundation in the mining industry of fossil fuels and in remote monitoring, helping to improve safety in the field [16].

E-learning, and especially the implementation of m-learning, is being facilitated by the IoE across the educational establishment, giving more accessibility to students. The benefits include more feedback and monitoring of the progress of the learners [17].

## IV. INTERNET OF NANO THINGS (IoNT)

The concept of IoE is being extended to its fullest by the implementation of the Internet of Nano Things (IoNT). This is achieved by incorporating nano-sensors in diverse objects and through the use of nano-networks. A model of this concept as a medical application is shown in Fig. 5: this provides access to data from places previously impossible to sense or from instruments inaccessible due to sensor size. This will enable new medical and environmental data to be collected, potentially leading to refinement of existing knowledge, new discoveries and better medical diagnostics [18]. The technology is described by Akyıldız and Josep Jornet using graphene-based nano-antennae operating at Terahertz frequencies. They also discuss the problems of extreme attenuation at these frequencies and networking at this nano-level [18].

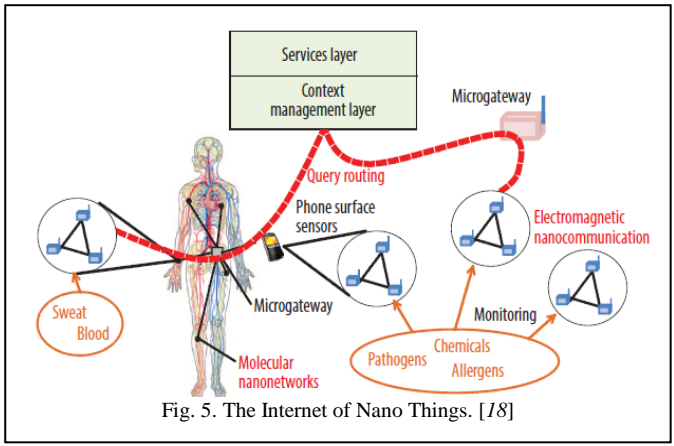

Fig. 5. The Internet of Nano Things. [18]

The interfacing of IoNT with existing micro-devices is important for it to become all-pervasive and further study should be focused on this task, especially in the industrial, biomedical and industrial arenas. Major challenges need to be addressed in the fields of electromagnetic channel modelling and the necessary networking protocols [19].

Each functional task, such as actuation or sensing, in an IoNT, is performed by a 'nano-machine' - whose dimensions may range from 1 to 100 nm [19].

## V. THE FUTURE INTERNET

Forbes [20] reported in August 2014 that the IoT had overtaken Big Data as a topic of discussion, as shown in the Gartner Hype Cycle [21] in Fig. 6, with over 45,000 references in the media in 2014, compared with only 15,000 in 2013.

The Gartner Hype Cycle shows the lifetime of a particular technology from inception to maturity to decline. This helps in business planning [21]. As Fig. 6 shows, the IoT is now at its peak above Big Data, with five to ten years to reach its maturity [20].

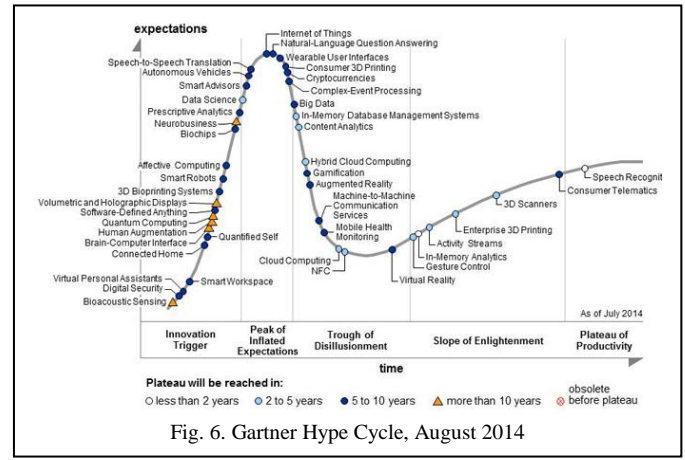

Fig. 6. Gartner Hype Cycle, August 2014

Much research is being conducted in the field of IoT in the three domains of the user experience, engineering and design [2]. The emphasis is particularly on the end-user and accessibility. This is especially pertinent as 50-200 billions artifacts are likely to be internetworked to the Internet by 2020 [2].

To help achieve a more user friendly interface, user-centred tools like Microsoft's Gadgeteer may be employed [2]. This tool provides rapid prototyping of connected devices [22]. Theories from cognitive psychology [23] have also been utilized to design adaptive IoT systems. This technique relies on using the "FRIEND::Process" tool for human task organization and for both bottom-up and top-down organizations [2].

Simpler embedded devices will form a significant part of the future IoT. Many difficult financial, technical and social issues remain to be addressed [22], but the reality is that the IoT does now exist and uses standardized international networking protocols [24] with IPv6 forming its core foundational routing protocol [25].

In order for the objects that compose the IoT to acquire 'ambient intelligence' they must comprehend the end user as completely as possible. This may be achieved by observing, monitoring and recording the human users': body movements, gestures, location, context and environment. This will be likely to lead to high levels of user support that were unknown previously in computing history [25]. The understanding of neuroscience, psychology and human behaviour will thus play an increasingly critical rôle in achieving device ambient intelligence. The devices must use Artificial Intelligence to understand how humans process information and interact appropriately within the right social context and multi-user scenarios [*23*].

The UK Open University offers users a course on IoT with programming and real-world sensing applications [*26*]. This is a first step in addressing the shortage of IoT engineers and programmers, especially as consumers become producers [*26*].

Educators will need to address many issues, not only the technical but also ethical and privacy issues. As a course, it was listed in the 2012 NMC Horizon Report [*27*]. The report also predicts IoT adoption around 2016-17.

Research continues with the European SENSEI project concentrating on the future underlying architecture of the IoT and its services [*24*].

For the IoT to be a practical pervasive reality, it must be able to coexist and integrate fully with the Cloud. This means using the current Internet Multimedia Subsystem (IMS) platform to integrate both technologies [*28*].

Due to the successful deployment of various novel, innovative and usefull applications based on IoT/IoE, the usage of computing devices and the Internet, by people from different cultures, socio-economic backgrounds, nations, religions and geographical diversity - is increasing at a near exponential rate. As a result of this phenomena, universal usability or Ubiquitous/Pervasive Computing [*29*,*30*], Usability [*31*], User Interface Design [*32*], specially Cross-Cultural Usability and Plasticity [*33*] of user interface design need to be focused. Exploring and analysing the Cross-Cultural Usability and Information System (IS) issues [*34*,*35*,*36*,*37*] focusing on web and mobile interaction using IoT/IoE as well as adoption trends and Diffusion of Innovations [*38*,*39*,*40*] need to be researched in depth. This is an important trend by users in how the 'IS' is utilized currently. As rightly pointed out by Ben Shneiderman, contemporary Computing is all about what users can do rather than what computers can do [*29*,*41*]**.** Thus for the future, the success of IoT/IoE must take into account the impact of cross-cultural usability by intensive research in this direction requiring further valuable research time.

## VI. Conclusion

As far back in 1984, the futurologist Ray Hammond, in his "The On-Line Handbook" [*42*], accurately foresaw that the linking of computers (i.e. the computer network and the Internet that we are using today) from all over the world would have far reaching effects, including: 1. The spread of knowledge; 2. The interchange of ideas and 3. The dissemination of information. Although he rightly further predicted that these were likely to bring a revolution in society, it is quite impossible to precisely point out where the current development in mobile applications, computer vision, consumer electronics, Artificial Intelligence and so forth will lead us. However, Henry Jenkins [*43*] has rightly explained the recent changes due to the digitization of media contents and their future impacts. We may experience a period of transition for novel interactions, ubiquitous computing, mobile and ambient intelligent applications and so forth in the remainder of the 21$^{st}$ Century, which has been observed for personal computers and other similar devices during the previous century. Although it cannot be guaranteed whether "Digital Immortality" as a form of "Technological Singularity" can be achieved by the year 2045 or not, as forecast by futurist Ray Kurzweil in his famous book "The Singularity is Near: When Humans Transcend Biology" [*44*] or whether "life [of "The World in 2030"] will be unrecognisable compared with life today" [*45*], our life is increasingly becoming digitalized with the inventions and adoptions of new technologies every day. Despite some negative aspects of this technological evolution, we can be optimistic about the coming computer revolution as technologies are becoming more affordable, convergent and novel in their solutions.

Understanding and interpreting these trends is strongly dependent on insights in classifying different aspects, such that links between those that are similar are clearly identified but differences between those that merit differentiation are also identified. In this connection, the separation between IoT, IoE and IoNT is seen to be a helpful differentiation that should aid insights in prediction of the near future.